\documentclass[10pt, conference, letterpaper]{IEEEtran}

\usepackage{amsmath,graphicx}
\usepackage{algorithm,algorithmic}
\usepackage{cite,amssymb,amsfonts,color,textcomp}
\usepackage{bm}
\usepackage{balance}
\usepackage{booktabs}
\usepackage{amsthm}
\usepackage{setspace}

\def\alphabf{\boldsymbol \alpha}
\def\betabf{\boldsymbol \beta }

\def\deltabf{\boldsymbol \delta }

\def\thetabf{\boldsymbol \theta}

\def\lambdabf{\boldsymbol \lambda }
\def\mubf{\boldsymbol \mu }

\def\phibf{\boldsymbol \phi }

\def\abf{{\bf a}}

\def\ebf{{\bf e}}

\def\hbf{{\bf h}}

\def\mbf{{\bf m}}
\def\nbf{{\bf n}}

\def\sbf{{\bf s}}

\def\vbf{{\bf v}}
\def\wbf{{\bf w}}
\def\xbf{{\bf x}}
\def\ybf{{\bf y}}
\def\zbf{{\bf z}}

\def\xbf{{\bf x}}
\def\ybf{{\bf y}}

\def\Ibf{{\bf I}}

\def\Ac{{\cal A}}
\def\Bc{{\cal B}}

\def\Fc{{\cal F}}

\def\Kc{{\cal K}}
\def\Lc{{\cal L}}

\def\Pc{{\cal P}}

\def\Sc{{\cal S}}
\def\Tc{{\cal T}}

\def\Yc{{\cal Y}}

\def\ie{{\it i.e.,\ \/}}
\def\nn{\nonumber}

\def\Re{\mathfrak{R}\mathfrak{e}}
\def\Im{\mathfrak{I}\mathfrak{m}}

\def\mae{{\mathbb{E}}}

\def\hhbf{{\hat{\hbf}}}

\newcommand{\olsi}[1]{\,\overline{\!{#1}}}

\theoremstyle{definition}

\newtheorem{assumption}{Assumption}
\newtheorem{proposition}{Proposition}

\newenvironment{mylist}%
{\begin{list}{}%
    {%
      \setlength{\itemindent}{-5pt}%
      \setlength{\leftmargin}{12pt}%
      \setlength{\parsep}{\parskip}
      \setlength{\labelsep}{5pt}
      \setlength{\itemsep}{2pt}}}%
  {\end{list}}

\begin{document}

\title{Robust Segmented Analog Broadcast Design to Accelerate Wireless Federated Learning}

\author{ Chong Zhang$^{\star}$, Ben Liang$^{\star}$, Min Dong$^{\dagger}$, Ali Afana$^{\ddagger}$, Yahia Ahmed$^{\ddagger}$\\
\normalsize $^{\star}$Department of Electrical and Computer Engineering, University of Toronto, Canada   \\
$^{\dagger}$Department of Electrical, Computer and Software Engineering, Ontario Tech University, Canada,
$^{\ddagger}$Ericsson Canada\thanks{This work was supported in part by Ericsson, the Natural Sciences and Engineering Research Council of Canada, and Mitacs.}
}%

\maketitle

\begin{abstract}
We consider downlink broadcast design for federated learning (FL) in a wireless network
with imperfect channel state information (CSI).
Aiming to reduce transmission latency,
we propose a segmented analog broadcast (SegAB) scheme,
where the parameter server, hosted by a multi-antenna base station, partitions the global model parameter vector
into segments and transmits multiple parameters from these segments simultaneously over a common downlink channel.
We formulate the SegAB transmission and reception processes
to characterize FL training convergence, capturing the effects of downlink beamforming
and imperfect CSI.
To maximize the FL training convergence rate, we establish an upper bound on the expected model
optimality gap and show that it can be minimized separately over the training rounds in online optimization, without requiring knowledge of the future channel states.
We solve the per-round problem to achieve robust downlink beamforming,
by minimizing the worst-case objective via
an epigraph representation and a feasibility subproblem
that ensures monotone convergence.
 Simulation with standard classification tasks under typical wireless network setting shows that the proposed SegAB substantially outperforms conventional
full-model per-parameter broadcast and other alternatives.
\end{abstract}

\section{Introduction}
\label{sec:intro}

\begin{figure*}[!htbp]
\centering
\hspace*{1.5em}\includegraphics[scale=.33]{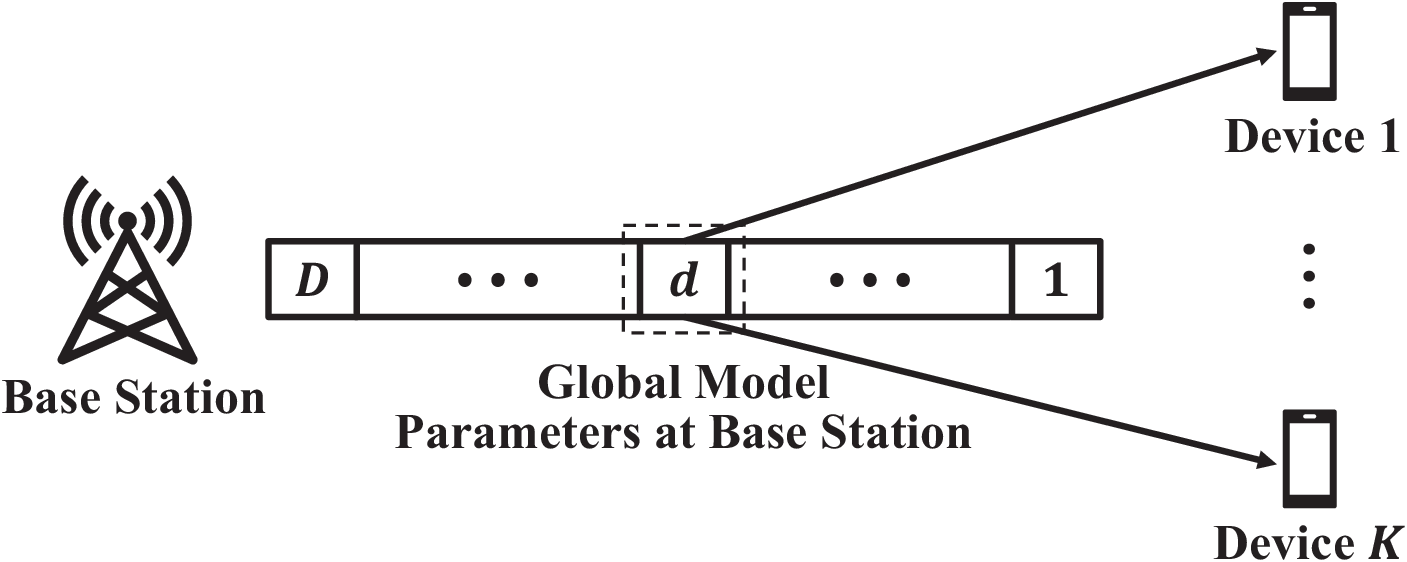}\hspace*{.2em}\includegraphics[scale=.33]{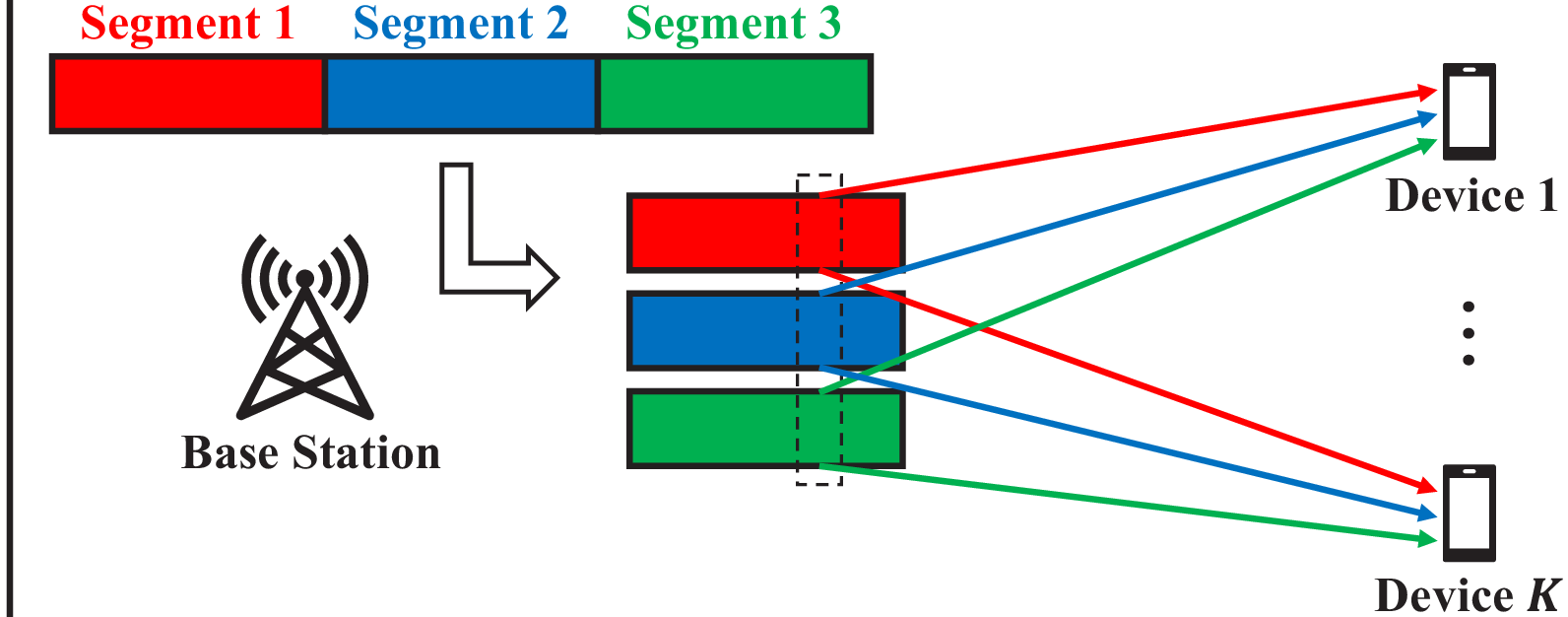}
\caption{Downlink analog broadcast for wireless FL. Left: traditional full-model per-parameter broadcast; Right: proposed SegAB
(an example of $S_t=3$ segments).}
\label{fig1:full_split_model_transmission}
\end{figure*}

Federated learning (FL) enables multiple devices to collaboratively train a machine learning (ML) model using their local data,
coordinated by a parameter server (PS) \cite{Mcmahan&etal:2017}. In wireless settings, the PS is typically hosted by
a base station (BS), which frequently exchanges large volumes of model parameters with devices over wireless links. This
stresses the limited communication resource, such as transmission bandwidth and power  \cite{Zhu&etal:2020}.
Furthermore, the signal distortion in wireless links leads to training errors that degrade the FL performance in training accuracy and convergence rate. Thus, efficient communication design is essential for wireless FL.

Most prior wireless FL works focused on improving the communication efficiency in uplink aggregation of local model parameters from the devices to the BS,
including digital transmit-then-aggregate schemes
such as
\cite{Du&etal:2020TSP,Chen&etal:jointLCFLTWC2021,Xu&Wang:2021TWC,Yang&etal:2021,Amiri&etal:TWC2021,Salehi&Hossain:TComm2021,Wang&etal:JSAC2022,Lan&etal:TWC2023}
and analog over-the-air computation
\cite{Zhu&etal:TWC2020,Zhang&Tao:TWC2021,Wang&etal:ToN2024,Yang&etal:TWC2020,Liu&etal:TWC2021,Kim&etal:TWC2023,Kalarde&etal:MSWiM2023,Kalarde&etal:2024}.
However, the downlink model broadcast is also critical for favorable FL learning performance.
Downlink transmission can be more vulnerable to noise and distortion than the uplink,
due to the challenge of maintaining reliable model broadcast to multiple users with varying channel conditions.
Downlink broadcast, if not properly designed, can become a bottleneck in terms of communication overhead, robustness,
and FL learning performance
\cite{Zheng&etal:2021JSAC,Amiri&etal:2020digitalDLarxiv,Amiri&etal:TWC2022,Guo&etal:JSAC2022,Shah&etal:2023JIOT,Wang&etal:JSAC2022b,Zhang&etal:WiOpt2023}.

Early works on downlink broadcast \cite{Zheng&etal:2021JSAC,Amiri&etal:2020digitalDLarxiv} studied classical digital transmission techniques.
However, it was later shown in \cite{Amiri&etal:TWC2022} that since the gradient descent in FL is noise
resilient, analog transmission can be more efficient.
Subsequently, several works studied downlink analog broadcast schemes
for single-antenna BSs with power control \cite{Guo&etal:JSAC2022,Shah&etal:2023JIOT,Wang&etal:JSAC2022b}
and multi-antenna BSs with transmit beamforming
\cite{Zhang&etal:WiOpt2023}.
However, these analog broadcast schemes all adopt the traditional full-model per-parameter transmission approach, as illustrated
in Fig.~\ref{fig1:full_split_model_transmission}-left,
where in each channel use, the BS sends one global model parameter to the devices.
However, due to the large number of model parameters in ML applications, this full-model approach can cause substantial delay and thus degraded FL performance.

To address this issue, we propose a Segmented Analog Broadcast (SegAB) framework for the downlink of FL,
which divides the  global model vector into multiple segments and simultaneously transmits different segments  to the devices.
As illustrated in Fig.~\ref{fig1:full_split_model_transmission}-right,
in each channel use of the downlink,
the BS broadcasts a weighted sum of parameters from all segments through multi-antenna beamforming.
By allowing simultaneous transmission of parameters, SegAB can substantially reduce the communication latency. However, it also introduces wireless interference among the different segments, which requires careful transmission design to balance the tradeoff between communication efficiency and broadcast accuracy, in order to optimize the overall FL performance.
Furthermore, analog transmission is susceptible to errors in the channel state information (CSI). Therefore, we require a robust design against imperfect CSI.

The main contribution of this paper is as follows:
\begin{mylist}
\item We propose a novel SegAB framework to allow simultaneous
transmission of global model parameters from multiple segments
in wireless FL.  SegAB only
requires a fraction of the channel uses that are needed
in the conventional full-model per-parameter transmission,
leading to substantially reduced transmission latency and faster FL training.
As far as we are aware, this is the first study on
segmented downlink broadcast in FL.

\item Aiming at optimizing the SegAB design to maximize
the FL training convergence rate, we formulate the SegAB
transmission and reception processes
and derive an overall model updating equation over each
communication round, capturing
 the impact of downlink
 beamforming and imperfect CSI on the global model update.
We then use the obtained model updating equation
to formulate a problem of
 robust downlink beamforming optimization.

\item For tractable design, we derive an upper bound on the expected
model optimality gap, which is separable over communication rounds with respect to beamforming optimization,
to allow efficient optimization design in an online fashion.
We show that each per-round subproblem admits a closed-form beamforming expression for robust optimization over imperfect CSI.
Using
an epigraph representation of the objective, we propose alternating updating
between the epigraph variable
and the beamforming vector via a feasibility subproblem.
This leads to an efficient robust downlink beamforming algorithm
with low-complexity closed-form beamforming updates and monotone convergence guarantee.

\item Simulation with standard classification tasks under typical wireless network setting
shows that SegAB substantially outperforms the conventional
approach of full-model per-parameter broadcast,
leading to faster training convergence for
a wide range of system configurations.
SegAB also outperforms other alternatives of beamforming for
segmented broadcast.
\end{mylist}

\section{Related Work}

Most analog schemes for wireless FL focus on the uplink aggregation of local model parameters,
in either single-antenna BS \cite{Zhu&etal:TWC2020,Zhang&Tao:TWC2021,Wang&etal:ToN2024}
or multi-antenna BS \cite{Yang&etal:TWC2020,Liu&etal:TWC2021,Kim&etal:TWC2023,Kalarde&etal:MSWiM2023,Kalarde&etal:2024}
variants.
These schemes leverage the superposition property of
the multiple access channel to achieve efficient
over-the-air aggregation for local model parameters.
They assume an error-free downlink to simplify design and analysis.

Other works have considered the downlink analog broadcast of global model
parameters \cite{Amiri&etal:TWC2022},
or joint downlink-uplink design
\cite{Guo&etal:JSAC2022,Shah&etal:2023JIOT,Wang&etal:JSAC2022b,Zhang&etal:WiOpt2023}.
These works demonstrate that well-designed analog broadcast
 schemes can significantly enhance the performance
  for wireless FL.
However, they all adopt the conventional full-model per-parameter transmission approach, which, as explained in Section~\ref{sec:intro},
can lead to severe inefficiency and degraded FL performance.

Recently, segmented transmission was
proposed for uplink over-the-air model aggregation
\cite{Zhang&etal:2025WiOpt}.
However, it is focused on analog aggregation among the devices, where each device transmits only a subset of the full model vector. Such uplink beamforming design is not applicable to our downlink broadcast problem.

Further recent works
have explored analog transmission for FL that
simultaneously trains multiple unrelated models (e.g., \cite{Zhang&etal:ICASSP2024}).
These works share similarity with ours as they also use beamforming to manage the interference in transmission. However, their optimization formulation and solution are essentially different, as their models are independent of each other, while the model segments in our work belong to and impact the performance of the same model.

\allowdisplaybreaks
\section{Wireless FL System Model}\label{sec:system_prob}

We consider a wireless FL system consisting of a server and $K$ devices.  Let  $\Kc = \{1, \ldots, K\}$ denote the set of devices. Each device $k \in \Kc$ holds a local training dataset, denoted by
$\Ac_{k} = \{(\abf_{k,i},y_{k,i}): 1 \le i \le A_k\}$, where
$\abf_{k,i}$ is the $i$-th data feature vector, and $y_{k,i}$ is the label for this data sample.
Using their respective local training datasets, the devices collaboratively
train a global ML model $\thetabf\in\mathbb{R}^{D\times 1}$ to predict the label of any feature vector, while keeping their local datasets private.
The local training loss function representing the training error at device $k$ is
\begin{align}
F_{k}(\thetabf)=\frac{1}{A_k}\sum_{i=1}^{A_k} L(\thetabf;\abf_{k,i},y_{k,i}) \nn
\end{align}
where $L(\cdot)$ is the sample-wise training loss function.
 The global training loss function is given by
\begin{align}
F(\thetabf) = \sum_{k=1}^{K}\frac{A_{k}}{A}F_{k}(\thetabf) \label{eq_global_local_equation}
\end{align}
where $A \triangleq  \sum_{k=1}^{K}A_{k}$.
The learning objective is to find an optimal global model $\thetabf^\star$ that minimizes $F(\thetabf)$.

We consider a standard FL procedure where the devices exchange model updates with the server via $T$ rounds of downlink-uplink communication.
Let $\Tc\triangleq\{0,\ldots,T-1\}$.
Round $t\in\Tc$ is given as follows:
\begin{mylist}
\item \emph{Downlink broadcast}: The server broadcasts the
$D$ parameters of the current global model $\thetabf_t$ to all $K$ devices.

\item  \emph{Local model update}: Device $k$ divides $\Ac_k$ into mini-batches, and applies the standard mini-batch stochastic gradient descent (SGD) algorithm over $J$ iterations to generate an updated local model
    based on $\thetabf_{t}$.
    Let $\thetabf^{\tau}_{k,t} $ be the local model update by device $k$ at iteration $\tau \in \{0,\ldots,J-1\}$, with
    $\thetabf^{0}_{k,t} = \thetabf_{t}$, and let  $\Bc^{\tau}_{k,t}$ denote the mini-batch at iteration  $\tau$.
    Then, the local model update is given by
    \begin{align}
    \thetabf^{\tau+1}_{k,t} & = \thetabf^{\tau}_{k,t} - \eta_t \nabla F_{k}(\thetabf^{\tau}_{k,t}; \Bc^{\tau}_{k,t})
    \label{eq_sgd}
    \end{align}
    where $\eta_{t}$ is the  learning rate.
    After  $J$ iterations, device $k$ obtains the updated local model $\thetabf^{J}_{k,t}$.

\item \emph{Uplink aggregation}: The devices send their updated local models $\{\thetabf^{J}_{k,t}\}_{k\in\Kc}$ to the server
through the uplink channel.
The server aggregates $\thetabf^{J}_{k,t}$'s to generate an updated global model $\thetabf_{t+1}$ for the next communication round $t+1$:
\begin{align}
\thetabf_{t+1} = \sum^{K}_{k=1}\frac{A_k}{A}\thetabf^{J}_{k,t}.
\label{eq_ul_update_OTA}
\end{align}

\end{mylist}

We consider wireless communication system where
the server is hosted by a BS equipped with $N$ antennas, and each device has a single antenna.
To efficiently utilize the communication bandwidth,
we adopt analog broadcast in the downlink
\cite{Amiri&etal:TWC2022,Guo&etal:JSAC2022,Shah&etal:2023JIOT,Wang&etal:JSAC2022b,Zhang&etal:WiOpt2023}. To focus on the downlink design in this work, we assume the uplink is error-free. As in \cite{Amiri&etal:TWC2022,Guo&etal:JSAC2022,Shah&etal:2023JIOT,Wang&etal:JSAC2022b,Zhang&etal:WiOpt2023},
we also assume that the system uses error-free digital transmission for control and signaling.
However, unlike previous works, we recognize that analog transmission is sensitive to CSI accuracy, so we consider a more practical scenario where the CSI estimation is imperfect. This calls for accurate tracking of the impact of CSI errors on model updating errors, and a careful design to improve the robustness of transmission and learning against imperfect CSI.

\section{Segmented Downlink Broadcast under Channel Estimation Error}\label{sec:FL_alg}

The proposed SegAB method has two salient features:
the simultaneous broadcast of different segments of the global model parameters, and robust downlink beamforming to reduce inter-segment interference.

\subsection{Segmented Analog Broadcast}\label{sec:SegAB_general}

The BS divides the model parameter vector into $S_t$ equal-sized segments,
with each segment having a length of $I_t \triangleq \lceil\frac{D}{S_t}\rceil$.
If $D$ is not a multiple of $S_t$,
the last segment will be padded with zero.
Let $\Sc_{t} \triangleq \{1, \ldots, S_{t}\}$ be the index set of model segments in communication round $t$.

The BS sends these $S_t$ segments simultaneously to the devices via the downlink channel,
shown in Fig.~\ref{fig1:full_split_model_transmission}-right.
Specifically, at the start of communication round $t$, the BS has the  current global model
$\thetabf_{t} = [\theta_{1,t}, \ldots, \theta_{D,t}]^{\textsf{T}}$.
Let $\sbf_{i,t}\in\mathbb{R}^{I_t\times 1}$ denote segment $i\in\Sc_{t}$ of $\thetabf_{t}$.
For efficient transmission,  we represent $\sbf_{i,t}$ using an equivalent complex vector $\tilde{\sbf}_{i,t}$,
whose real and imaginary parts contain the first and second halves of the elements in $\sbf_{i,t}$, respectively.
That is, $\tilde{\sbf}_{i,t} = \tilde{\sbf}^{\text{re}}_{i,t} + j\tilde{\sbf}^{\text{im}}_{i,t}\in\mathbb{C}^{\frac{I_t}{2}\times 1}$, where $\tilde{\sbf}^{\text{re}}_{i,t}$ contains the first
$\frac{I_t}{2}$ elements in $\sbf_{i,t}$ and $\tilde{\sbf}^{\text{im}}_{i,t}$ contains the other $\frac{I_t}{2}$ elements.

Let $\hbf_{k,t}\in\mathbb{C}^{N\times 1}$ be the downlink channel vector from the BS to device $k\in\Kc$ in communication round $t$.
The BS employs transmit beamforming to send the $S_t$ complex global model segments $\tilde{\sbf}_{i,t}$'s
simultaneously to the $K$ devices.
Specifically,
let $\wbf_{i,t}\in\mathbb{C}^{N\times 1}$
be the BS transmit beamforming vector for
model segment $i$ in communication round $t$.
The BS uses $\wbf_{i,t}$ to transmit
the normalized complex global model segment
$\frac{\tilde{\sbf}_{i,t}}{u_{i,t}}$,
with $u_{i,t} \triangleq \frac{\|\tilde{\sbf}_{i,t}\|}{\sqrt{I_t/2}}$.
Thus, the received signal vector $\zbf_{k,t}\in\mathbb{C}^{\frac{I_t}{2}\times 1}$ at device $k$
over the $\frac{I_t}{2}$ channel uses is given by
\begin{align}
\zbf_{k,t} = \sum^{S_t}_{i=1}\wbf^\textsf{H}_{i,t}\hbf_{k,t}\frac{\tilde{\sbf}_{i,t}}{u_{i,t}} + \nbf_{k,t}  \nn
\end{align}
where
$\nbf_{k,t}\sim \mathcal{CN}({\bf 0}, \sigma^2\Ibf)$ is  the receiver additive white Gaussian noise (AWGN) vector.

The transmit beamforming vectors $\{\wbf_{i,t}\}_{i=1}^{S_t}$ are subject to the BS transmit power budget.
Let $P$ be the average transmit power limit per channel use
at the BS.
Then, we have the BS transmit power constraint $\sum^{S_t}_{i=1}\|\wbf_{i,t}\|^2 \leq P$.

\subsection{Receiver Processing by Devices}\label{sec:receiver_process_dev}

The BS also sends the CSI to device $k\in\Kc$ via the downlink signaling channel
to facilitate receiver processing.
In practice, the BS has access only to estimated channel
vectors.
Let $\hat{\hbf}_{k,t}$ denote the estimated channel at the BS for device $k$ in communication round $t$.
The estimation error $\Delta\hbf_{k,t}$ in the channel vector is modeled as
 $\hbf_{k,t} = \hat{\hbf}_{k,t} + \Delta\hbf_{k,t}$.
We assume that the channel estimation error $\Delta\hbf_{k,t}$ is bounded \cite{Godara2001handbook,Shenouda&Davidson:2007JSTSP}:
\begin{align}
\|\Delta\hbf_{k,t}\| \leq \epsilon_{k,t}, \quad \forall k\in\Kc, \forall t.
\label{eq_channel_error_bound}
\end{align}

Based on estimated channel $\hat{\hbf}_{k,t}$, the BS sends $S_t$ scaling factors, in the form of
$u_{i,t}\frac{\hhbf^\textsf{H}_{k,t}\wbf_{i,t}}{|\hhbf^\textsf{H}_{k,t}\wbf_{i,t}|^2}$ for $i\in\Sc_t$,
to device $k$ via the downlink signaling channel for the device to recover each segment $i\in\Sc_{t}$ from $\zbf_{k,t}$.
Then, for segment $i$, device $k$ post-processes $\zbf_{k,t}$
using the corresponding scaling factor,
leading to
\begin{align}
\hat{\tilde{\sbf}}^{i}_{k,t} & = u_{i,t}\frac{\hhbf^\textsf{H}_{k,t}\wbf_{i,t}}{|\hhbf^\textsf{H}_{k,t}\wbf_{i,t}|^2}\zbf_{k,t} = \frac{\hhbf^\textsf{H}_{k,t}\wbf_{i,t}\wbf^\textsf{H}_{i,t}\hbf_{k,t}}{|\hhbf^\textsf{H}_{k,t}\wbf_{i,t}|^2}\tilde{\sbf}_{i,t}\nn\\
& + \sum_{j\neq i}\frac{\hhbf^\textsf{H}_{k,t}\wbf_{i,t}\wbf^\textsf{H}_{j,t}\hbf_{k,t}}{|\hhbf^\textsf{H}_{k,t}\wbf_{i,t}|^2}\cdot
\frac{u_{i,t}\tilde{\sbf}_{j,t}}{u_{j,t}} + \tilde{\nbf}^{i}_{k,t}
\label{eq_process_signal_p1}
\end{align}
where $\tilde{\nbf}^{i}_{k,t} \triangleq u_{i,t}\frac{\hhbf^\textsf{H}_{k,t}\wbf_{i,t}}{|\hhbf^\textsf{H}_{k,t}\wbf_{i,t}|^2}\nbf_{k,t}$
is the post-processed noise vector at device $k$ for segment $i$.
Then, device $k$ obtains the estimate of segment $\sbf_{i,t}$ as
$\hat{\sbf}^{i}_{k,t} = [\Re{\{\hat{\tilde{\sbf}}^{i}_{k,t}\}^\textsf{T}}, \Im{\big\{\hat{\tilde{\sbf}}^{i}_{k,t}\}^\textsf{T}}]^\textsf{T}$.

Finally, device $k$ stacks all segment estimates
$\{\hat{\sbf}^{i}_{k,t}\}_{i\in\Sc_{t}}$ to obtain
an estimate for the global model $\thetabf_{t}$, given by
$\hat{\thetabf}_{k,t}= [(\hat{\sbf}^{1}_{k,t})^\textsf{T}, \ldots, (\hat{\sbf}^{S_t}_{k,t})^\textsf{T}]^\textsf{T}$.

\subsection{Robust Downlink Beamforming Problem Formulation}\label{sec:prob_form}

Aiming to maximize the FL training convergence rate
in the presence of channel uncertainty,
we consider a robust design of the downlink
beamforming, where we optimize transmit beamforming vectors
$\{\wbf_{i,t}\}$ to maximize the worst-case performance
over channel estimation errors $\{\Delta\hbf_{k,t}\}$.
Specifically, after device $k$ obtains its global model estimate $\hat{\thetabf}_{k,t}$
as described in Section~\ref{sec:receiver_process_dev},
it is used as an initial point for the local training \eqref{eq_sgd},
\ie $\thetabf^{0}_{k,t} = \hat{\thetabf}_{k,t}$.
After $J$ mini-batch SGD iterations of local training,
the local model update at device $k$ is given by
\begin{align}
\thetabf^{J}_{k,t} = \hat{\thetabf}_{k,t} + \Delta\thetabf_{k,t}
\label{eq_real_local_model_update}\vspace{-1em}
\end{align}
where $\Delta\thetabf_{k,t} \triangleq - \eta_t\sum^{J-1}_{\tau=0}\nabla F_{k}(\thetabf^{\tau}_{k,t}; \Bc^{\tau}_{k,t})$ denotes
the local model change at device $k$ after the local training.
Combining \eqref{eq_real_local_model_update} and \eqref{eq_ul_update_OTA}, we obtain the
global model update $\thetabf_{t+1}$ for the next communication round $t+1$ as\\[-.1em]
\begin{align}
\thetabf_{t+1} = \sum^{K}_{k=1}\frac{A_k}{A}\hat{\thetabf}_{k,t} + \sum^{K}_{k=1}\frac{A_k}{A}\Delta\thetabf_{k,t}.
\label{eq_ul_update_OTA_final}
\end{align}
The global model updating equation
\eqref{eq_ul_update_OTA_final}
shows the impact of the segmented downlink broadcast processing
and the local model updates, represented by
the first and second terms at the right-hand side.

Let $\wbf_t\triangleq[\wbf_{1,t}^{\textsf{H}},
\ldots, \wbf_{S_t,t}^{\textsf{H}}]^{\textsf{H}}\in\mathbb{C}^{S_tN\times 1}$
denote the stacked BS transmit beamforming vectors in communication round $t$.
We design robust downlink transmit beamforming
to minimize the worst-case expected model optimality gap to $\thetabf^\star$ over the channel estimation errors
after $T$ communication rounds, formulated as
\begin{align}
\Pc_{o}: \, \min_{\{ \wbf_t\}^{T-1}_{t=0}} & \, \max_{ \substack{\|\Delta\hbf_{k,t}\|\leq\epsilon_{k,t}, \\ k\in\Kc,t\in\Tc} } \, \mae[\|\thetabf_{T}- \thetabf^\star\|^2] \nn\\
\text{s.t.} \ \ &  \|\wbf_{t}\|^2 \leq P, \quad t\in\Tc
\label{constra_BS_transmit_power}
\end{align}
where $\mae[\cdot]$ is taken over the receiver noise
at the devices and mini-batch sampling for local training,
and constraint \eqref{constra_BS_transmit_power} is the
BS transmit power constraint.
Note that since the communication time of
each round is fixed for a given model size $D$
and number of segments $S_t$,
problem $\Pc_{o}$ also represents
minimizing the worst-case optimality
gap after training for some certain wall-clock time.

There are several challenges in solving problem $\Pc_{o}$.
First, $\|\Delta\hbf_{k,t}\|=\epsilon_{k,t}$ in general
is not the worst case for the maximization of $\mae[\|\thetabf_{T}- \thetabf^\star\|^2]$, so the inner maximization of problem $\Pc_{o}$ is nontrivial.
Furthermore, in FL training it is generally
impossible to directly evaluate the expected
model optimality gap at round $T$.
Finally, problem $\Pc_{o}$ is a finite-horizon stochastic optimization problem for $T$ rounds, which necessitates an online solution since future channel states during FL training are unavailable.

\section{Robust Downlink Beamforming Optimization}
\label{sec:upper_bound}

To solve problem $\Pc_{o}$, we first develop a more tractable upper bound on $\mae[\|\thetabf_{T}- \thetabf^\star\|^2]$
by analyzing the convergence rate of the global model update,
as a function of the beamforming vectors $\{\wbf_t\}_{t\in\Tc}$
and channel estimation errors $\{\Delta\hbf_{k,t}\}_{k\in\Kc,t\in\Tc}$.
Then, observing the per-round separable structure of this upper bound, we propose a robust downlink beamforming method to minimize it over worst-case channel estimation errors.

\subsection{Convergence Analysis of FL Training with SegAB}
\label{sec:converg_analysis_FL_SegAB}
To analyze the FL convergence rate with SegAB, we make the following assumptions.
They are commonly used in the existing literature for FL convergence analysis,
and here we extend them to SegAB.

\begin{assumption}\label{assump_smooth}
The local loss functions $F_k(\cdot)$'s are differentiable and are  $L$-smooth:
For $\forall \xbf, \ybf\in\mathbb{R}^{D\times 1}$,
$F_k(\ybf) \leq F_k(\xbf) + (\ybf - \xbf)^{\textsf{T}}\nabla F_k(\xbf) + \frac{L}{2}\|\ybf - \xbf\|^2$,
$\forall k\in\Kc$.
Furthermore,
$F_k(\cdot)$'s are  $\lambda$-strongly convex:
For $\forall \xbf, \ybf\in\mathbb{R}^{D\times 1}$,
$F_k(\ybf) \geq F_k(\xbf) + (\ybf - \xbf)^{\textsf{T}}\nabla F_k(\xbf) + \frac{\lambda}{2}\|\ybf - \xbf\|^2$, $\forall k\in\Kc$.
\end{assumption}

Let $\nabla F$ denote the gradient of the global loss function given in
\eqref{eq_global_local_equation}.
Denote the part of gradients $\nabla F(\thetabf_{t})$ and
$\nabla F_{k}(\thetabf^{\tau}_{k,t})$,
corresponding to segments $i$, by $\nabla F^{i}(\thetabf_{t})$ and $\nabla F^i_{k}(\thetabf^{\tau}_{k,t})$,
respectively.

\begin{assumption}\label{assump_bound_diff}
The deviations among the gradients of the global and local loss functions are  bounded:
For some $\phi_i \ge 0$,
$\mae[\| \nabla F^{i}(\thetabf_{t}) - \sum^{K}_{k=1}c^i_{k}
\nabla F^i_{k}(\thetabf^{\tau}_{k,t})  \|^2] \leq \phi_{i}$,
where $c^i_{k} \geq 0$,
 $\sum^{K}_{k=1}c^i_{k}=1$,
$\forall i\in\Sc_t$, $\forall \tau$, $\forall t$.
For some $\zeta_{i} \ge 0$,
$\mae[\| \nabla F^i_{k}(\thetabf^{\tau}_{k,t}) -
\nabla F^i_{k}(\thetabf^{\tau}_{k,t}; \Bc^{\tau}_{k,t})\|^2]
\leq \zeta_{i}$,
$\forall i\in\Sc_t$, $\forall k\in\Kc$, $\forall \tau$, $\forall t$.
\end{assumption}

We apply the channel estimation error model
$\hbf_{k,t} = \hat{\hbf}_{k,t} + \Delta\hbf_{k,t}$
and $\frac{u_{i,t}}{u_{j,t}} = \frac{\|\tilde{\sbf}_{i,t}\|}{\|\tilde{\sbf}_{j,t}\|}$
to \eqref{eq_process_signal_p1}.
Let
\begin{align}
&\tilde{\ebf}^{i}_{k,t} \triangleq \frac{\hhbf^\textsf{H}_{k,t}\wbf_{i,t}\wbf^\textsf{H}_{i,t}\Delta\hbf_{k,t}}{|\hhbf^\textsf{H}_{k,t}\wbf_{i,t}|^2}\tilde{\sbf}_{i,t}
\nn\\
&+ \sum_{j\neq i}\frac{\hhbf^\textsf{H}_{k,t}\wbf_{i,t}\wbf^\textsf{H}_{j,t}(\hhbf_{k,t} + \Delta\hbf_{k,t})}{|\hhbf^\textsf{H}_{k,t}\wbf_{i,t}|^2}\cdot
\frac{\|\tilde{\sbf}_{i,t}\|\tilde{\sbf}_{j,t}}{\|\tilde{\sbf}_{j,t}\|} +  \tilde{\nbf}^{i}_{k,t}.
\label{eq_error_equa_accu}
\end{align}
Let $\ebf^{i}_{k,t} = [\Re{\{\tilde{\ebf}^{i}_{k,t}\}^\textsf{T}}, \Im{\{\tilde{\ebf}^{i}_{k,t}\}^\textsf{T}}]^\textsf{T}$
and stack $\ebf^{i}_{k,t}$, $i\in\Sc_t$
as $\ebf_{k,t} \triangleq [(\ebf^{1}_{k,t})^\textsf{T}, \ldots, (\ebf^{S_t}_{k,t})^\textsf{T}]^\textsf{T}$.
Based on \eqref{eq_process_signal_p1} and \eqref{eq_error_equa_accu},
we have $\hat{\thetabf}_{k,t} = \thetabf_{t} + \ebf_{k,t}$.
We then rewrite \eqref{eq_ul_update_OTA_final} as
\begin{align}
\thetabf_{t+1} - \thetabf^\star = \thetabf_{t} - \thetabf^\star + \sum^{K}_{k=1}\frac{A_k}{A}\Delta\thetabf_{k,t} + \sum^{K}_{k=1}\frac{A_k}{A}\ebf_{k,t}.
\label{eq_global_model_update_final}
\end{align}
The term
$\ebf_{k,t}$ represents the accumulated transmission error
at device $k$ in communication round $t$.
It accounts for the impact of the channel estimation error,
inter-segment interference, and device receiver noise.

Based on Assumptions~\ref{assump_smooth} and \ref{assump_bound_diff}, and \eqref{eq_global_model_update_final},
we provide an upper bound on the expected model optimality gap $\mae[\|\thetabf_{T}- \thetabf^\star\|^2]$
after $T$ communication rounds, capturing the inter-segment interference and transmit beamforming processing under channel estimation error.

\begin{proposition}\label{thm:convergence}
Let $\nu \triangleq \max_{i,t} \|\sbf_{i,t}\|^2$.
For SegAB described in Section~\ref{sec:FL_alg},
under Assumptions~\ref{assump_smooth} and \ref{assump_bound_diff} and for
$\eta_t<\frac{1}{L}$, $\forall t\in\Tc$, the expected model optimality gap $\mae[\|\thetabf_{T}- \thetabf^\star\|^2]$,
after $T$ communication rounds is upper bounded by
\begin{align}
& \mae[\|\thetabf_{T} - \thetabf^\star\|^2] \leq\sum^{T-1}_{t=0}\olsi{G}_{t}( H_t(\wbf_t,\Delta\hbf_{t}) +  C_{t} )  +\Gamma\prod_{t=0}^{T-1}G_t
\label{eq_thm1}
\end{align}
where $\Gamma \triangleq \| \thetabf_{0} - \thetabf^\star\|^2$,  $G_t \triangleq 4(1-\eta_t\lambda)^{2J}$,
 $C_{t} \triangleq 4\eta^2_{t}J^2\sum^{S_t}_{i=1}(\phi_{i} + \zeta_{i})$,
$\olsi{G}_{t} \triangleq \prod^{T-1}_{s=t+1}G_{s}$
with $\olsi{G}_{T-1}=1$,
$\Delta\hbf_{t}\triangleq[\Delta\hbf_{1,t}^{\textsf{H}},
\ldots, \Delta\hbf_{K,t}^{\textsf{H}}]^{\textsf{H}}$,
and
\begin{align}
& H_t(\wbf_t,\Delta\hbf_{t})  \triangleq
16S_t\nu\sum^{S_t}_{i=1}\sum^{K}_{k=1}\frac{A_k}{A}\Bigg(\frac{ \sum^{S_t}_{j=1} |\Delta\hbf^\textsf{H}_{k,t}\wbf_{j,t}|^2 }{ |\hhbf^\textsf{H}_{k,t}\wbf_{i,t}|^2  }  \nn\\
&\qquad\qquad\qquad\qquad + \frac{ \sum_{j\neq i}|\hhbf^\textsf{H}_{k,t}\wbf_{j,t}|^2 + \sigma^2 }{ |\hhbf^\textsf{H}_{k,t}\wbf_{i,t}|^2  }\Bigg). \label{eq_H_func}
\end{align}
\end{proposition}
\IEEEproof
See Appendix~\ref{appA}.
\endIEEEproof

A key property of this upper bound is that it
is separable across communication rounds, leading to
efficient optimization design in an online fashion.
Below, we leverage this property and propose a
low-complexity solution to robust downlink beamforming design
with fast closed-form updates.

\subsection{Robust Downlink Beamforming Design}
\label{sec:joint_bf}

\subsubsection{Conversion to Per-round Problem}
We replace the objective function in $\Pc_{o}$
with the upper bound in \eqref{eq_thm1}.
Omitting the constant terms in \eqref{eq_thm1} that do not depend on $\{\wbf_t\}$ or $\{\Delta\hbf_{t}\}$,
we obtain the following optimization problem:
\begin{align}
\Pc_{1}: \, \min_{\{ \wbf_t\}^{T-1}_{t=0}} & \, \max_{ \substack{\|\Delta\hbf_{k,t}\|\leq\epsilon_{k,t}, \\ k\in\Kc,t\in\Tc} } \, \sum^{T-1}_{t=0}\olsi{G}_{t}H_t(\wbf_t,\Delta\hbf_{t}) \nn\\
\text{s.t.} \ \ &  \|\wbf_{t}\|^2 \leq P, \quad t\in\Tc. \nn
\end{align}
Following Proposition~\ref{thm:convergence} and Assumption~\ref{assump_smooth},
we set $\eta_t < \frac{1}{L} \leq \frac{1}{\lambda}$, $\forall t\in\Tc$.
This leads to $G_t > 0$ and further $\olsi{G}_{t} > 0$.
Thus, problem $\Pc_{1}$ can be equivalently decomposed
into $T$ per-round problems.
The problem at round $t$ is given by
\begin{align}
\Pc_{2,t}: \, \min_{ \wbf_t } & \, \max_{ \substack{\|\Delta\hbf_{k,t}\|\leq\epsilon_{k,t}, \\ k\in\Kc} } \, H_t(\wbf_t,\Delta\hbf_{t}) \nn\\
\text{s.t.} \ \ &  \|\wbf_{t}\|^2 \leq P. \nn
\end{align}

\subsubsection{Inner Maximization}
Problem $\Pc_{2,t}$ is a min-max optimization problem, which is challenging to solve in general.
However, we can show that the inner maximization $\max_{ \|\Delta\hbf_{k,t}\|\leq\epsilon_{k,t}, k\in\Kc } H_t(\wbf_t,\Delta\hbf_{t})$
in $\Pc_{2,t}$ has a closed-form solution $\Delta\hbf_{t}$ that depends on $\wbf_t$.
Specifically, it can be decomposed into $K$ equivalent subproblems,
each for a device $k\in\Kc$, given by
\begin{align}
\Pc^{\text{max}}_{2,k,t}: \,
\max_{ \|\Delta\hbf_{k,t}\|\leq\epsilon_{k,t} } \, \sum^{S_t}_{i=1} |\Delta\hbf^\textsf{H}_{k,t}\wbf_{i,t}|^2. \nn
\end{align}
This is an eigenvalue problem.
Let $\phibf_{k,t}(\wbf_{t})$ denote
the unit-norm eigenvector corresponding to the
largest eigenvalue of its objective function.
The solution to $\Pc^{\text{max}}_{2,k,t}$ is
\begin{align}
\Delta\hbf_{k,t} = \epsilon_{k,t}\phibf_{k,t}(\wbf_{t}).  \label{eq_channel_error_solution}
\end{align}
Based on \eqref{eq_channel_error_solution}, we obtain the worst-case objective function $H_t(\wbf_t,\{\epsilon_{k,t}\phibf_{k,t}(\wbf_{t})\})$
over channel estimation errors.
We then minimize this expression to obtain a solution $\wbf_t$ for robust beamforming design as follows.

\subsubsection{Per-round Beamforming Optimization}
Given the worst-case channel estimation error $\Delta\hbf_{t}$ in \eqref{eq_channel_error_solution},
we can rewrite problem $\Pc_{2,t}$ as
\begin{align}
\Pc^{1\text{min}}_{2,t}: \, &\min_{ \wbf_t } \,\, H_t(\wbf_t,\{\epsilon_{k,t}\phibf_{k,t}(\wbf_{t})\}) \nn\\
& \text{s.t.} \ \ \|\wbf_{t}\|^2 \leq P. \nn
\end{align}
Based on the expression of $H_t(\wbf_t,\Delta\hbf_{t})$
in \eqref{eq_H_func}, we transform $\Pc^{1\text{min}}_{2,t}$ into
an equivalent joint optimization problem
using the epigraph representation:
\begin{align}
& \Pc^{2\text{min}}_{2,t}: \, \min_{\wbf_t,\mubf_{t}} \, {\bf{1}}^{\textsf{T}}\mubf_{t}  \nn\\
& \text{s.t.} \ \ r_{k}\epsilon^2_{k,t}\sum^{S_t}_{j=1}|\phibf^{\textsf{H}}_{k,t}(\wbf_t)\wbf_{j,t}|^2  + r_{k}\sum_{j\neq i}|\hhbf^\textsf{H}_{k,t}\wbf_{j,t}|^2 + \sigma_{k}^2  \nn\\
& \qquad\qquad\qquad\qquad  \leq \mu^{i}_{k,t} |\hhbf^\textsf{H}_{k,t}\wbf_{i,t}|^2, \,\, k\in\Kc,i\in\Sc_t, \nn\\
& \quad\,\,\,\, \|\wbf_t\|^2\leq P  \nn
\end{align}
where we introduce new vector $\mubf_t \triangleq [\mu^{1}_{1,t}, \ldots, \mu^{S_t}_{K,t}]^{\textsf{T}}\in\mathbb{R}^{S_tK\times 1}$,
and $r_{k} \triangleq \frac{A_k}{A}$,
and $\sigma_{k} \triangleq \sqrt{ r_k }\sigma$.
Since the problem is nonconvex, we propose to alternatingly optimize
the BS transmit beamformer $\wbf_t$
and the epigraph variable $\mubf_{t}$.
The two subproblems are given below:

{\it{i) Updating $\wbf_{t}$:}}
\label{sec:w_update}
Given $\mubf_{t}$, we can transform $\Pc^{2\text{min}}_{2,t}$
into a feasibility problem, given by
\begin{align}
& \Pc^{1\text{fea}}_{2,t}: \,  {\text{Find}} \,\, \{\wbf_t\}  \nn\\
& \text{s.t.} \ \ r_{k}\epsilon^2_{k,t}\sum^{S_t}_{j=1}|\phibf^{\textsf{H}}_{k,t}(\wbf_t)\wbf_{j,t}|^2  + r_{k}\sum_{j\neq i}|\hhbf^\textsf{H}_{k,t}\wbf_{j,t}|^2 + \sigma_{k}^2  \nn\\
& \qquad\qquad\qquad\qquad  \leq \mu^{i}_{k,t} |\hhbf^\textsf{H}_{k,t}\wbf_{i,t}|^2, \,\, k\in\Kc,i\in\Sc_t, \nn\\
& \quad\,\,\,\, \|\wbf_t\|^2\leq P.  \nn
\end{align}

{\it{ii) Updating $\mubf_{t}$:}}
\label{sec:tau_update}
Given $\wbf_t$, the optimal $\mu^{i}_{k,t}$, $k\in\Kc,i\in\Sc_t$, is given by
\begin{align}
\mu^{i}_{k,t}\! =\! \frac{ r_{k}\epsilon^2_{k,t}\!\sum^{S_t}_{j=1}\!|\phibf^{\textsf{H}}_{k,t}(\wbf_t)\wbf_{j,t}|^2\!\!  +\! r_{k}\!\sum_{j\neq i}\!|\hhbf^\textsf{H}_{k,t}\wbf_{j,t}|^2\!\! +\! \sigma_{k}^2 }{|\hhbf^\textsf{H}_{k,t}\wbf_{i,t}|^2}.
\label{Update_tau_AO}
\end{align}

Note that directly solving the feasibility problem
$\Pc^{1\text{fea}}_{2,t}$ will not be effective in  alternating optimization, since finding a feasible solution to $\Pc^{1\text{fea}}_{2,t}$
cannot guarantee a descending objective value ${\bf{1}}^{\textsf{T}}\mubf_{t}$
over iterations.
To find an efficient solution $\wbf_t$
that leads to a strictly decreasing objective value,
we propose to first modify $\Pc^{1\text{fea}}_{2,t}$
to the following, by introducing a slack variable
$\deltabf_{t} \triangleq [\delta^{1}_{1,t}, \ldots, \delta^{S_t}_{K,t}]^{\textsf{T}}\in\mathbb{R}^{S_tK\times 1}$:
\begin{align}
& \Pc^{2\text{fea}}_{2,t}: \, \max_{\deltabf_{t},\wbf_t} \, {\bf{1}}^{\textsf{T}}\deltabf_{t}  \nn\\
& \text{s.t.} \ \ r_{k}\epsilon^2_{k,t}\sum^{S_t}_{j=1}|\phibf^{\textsf{H}}_{k,t}(\wbf_t)\wbf_{j,t}|^2 + r_{k}\sum_{j\neq i}|\hhbf^\textsf{H}_{k,t}\wbf_{j,t}|^2
+ \delta^{i}_{k,t}  \nn\\
& \qquad\qquad\qquad  + \sigma_{k}^2 \leq \mu^{i}_{k,t} |\hhbf^\textsf{H}_{k,t}\wbf_{i,t}|^2, \,\, k\in\Kc,i\in\Sc_t, \label{constra_sinr_3}\\
& \quad\,\,\,\, \|\wbf_t\|^2\leq P.  \nn
\end{align}
Furthermore, we consider an upper bound on
the left-hand side of constraint \eqref{constra_sinr_3}.
Specifically, by the Cauchy-Schwarz inequality and
$\|\wbf_t\|^2\leq P$, we obtain
\begin{align}
\sum^{S_t}_{j=1}|\phibf^{\textsf{H}}_{k,t}(\wbf_t)\wbf_{j,t}|^2 \leq \sum^{S_t}_{j=1}\|\wbf_{j,t}\|^2 \leq P.   \label{eq_cs_power}
\end{align}
We replace the corresponding expression in \eqref{constra_sinr_3}
by its upper bound, arriving at
\begin{align}
& \Pc^{3\text{fea}}_{2,t}: \, \max_{\deltabf_{t},\wbf_t} \, {\bf{1}}^{\textsf{T}}\deltabf_{t}  \nn\\
& \text{s.t.} \ \ r_{k}\sum_{j\neq i}|\hhbf^\textsf{H}_{k,t}\wbf_{j,t}|^2 + \delta^{i}_{k,t} + r_{k}\epsilon^2_{k,t}P + \sigma_{k}^2 \nn\\
& \qquad\qquad\qquad\qquad  \leq \mu^{i}_{k,t} |\hhbf^\textsf{H}_{k,t}\wbf_{i,t}|^2, \,\, k\in\Kc,i\in\Sc_t, \label{constra_sinr_4}\\
& \quad\,\,\,\, \|\wbf_t\|^2\leq P.  \nn
\end{align}
Note that the benefit of the above optimization approach is
two-fold. First, the constant bound $P$ replaces the term
$\sum^{S_t}_{j=1}|\phibf^{\textsf{H}}_{k,t}(\wbf_t)\wbf_{j,t}|^2$,
which contains an implicit function of $\wbf_t$
that makes it hard to solve $\Pc^{2\text{fea}}_{2,t}$.
Second, using the solution $\wbf_t$ from $\Pc^{3\text{fea}}_{2,t}$
to update $\mu^{i}_{k,t}$ will lead to a smaller value of
 $\mu^{i}_{k,t}$ than in the previous iteration.
We also note that a solution
to $\Pc^{3\text{fea}}_{2,t}$
is also a feasible solution to $\Pc^{1\text{fea}}_{2,t}$.
Below, we focus on solving $\Pc^{3\text{fea}}_{2,t}$
efficiently via a low-complexity algorithm.

Problem $\Pc^{3\text{fea}}_{2,t}$ is still a nonconvex problem.
We employ the successive convex approximation (SCA) method \cite{Marks&Wright:OperReas1978}
to solve it.
Given any auxiliary vector $\vbf_{i}\in \mathbb{C}^{N\times 1}$, $i\in\Sc_t$,
since $(\wbf_{i,t} - \vbf_{i})^{\textsf{H}}\hhbf_{k,t}
\hhbf_{k,t}^{\textsf{H}}(\wbf_{i,t} - \vbf_{i}) \ge 0$,
we have $|\hhbf^{\textsf{H}}_{k,t}\wbf_{i,t}|^{2} \ge
2\Re{\{\vbf^{\textsf{H}}_{i}\hhbf_{k,t}\hhbf_{k,t}^{\textsf{H}}\wbf_{i,t}\}}-|\hhbf^{\textsf{H}}_{k,t}\vbf_i|^{2}$.
Let $\vbf\triangleq [\vbf^{\textsf{H}}_{1},\ldots,\vbf^{\textsf{H}}_{S_t}]^{\textsf{H}}\in\mathbb{C}^{S_tN\times 1}$.
We use the above bound to convexify
$\Pc^{3\text{fea}}_{2,t}$
into a sequence of subproblems,
which take $\vbf$ as parameter and have the following form:
\begin{align}
& \Pc^{\text{sca}}_{2,t}(\vbf): \, \min_{\deltabf_{t},\wbf_t} \, -{\bf{1}}^{\textsf{T}}\deltabf_{t}  \nn\\
& \text{s.t.} \ r_{k}\!\!\sum_{j\neq i}\!|\hhbf^\textsf{H}_{k,t}\!\wbf_{j,t}|^2\!\! -\! 2\mu^{i}_{k,t}\Re{\{\!\vbf^{\textsf{H}}_{i}\hhbf_{k,t}\hhbf_{k,t}^{\textsf{H}}\!\wbf_{i,t}\!\}}\! +\! \mu^{i}_{k,t}|\hhbf^{\textsf{H}}_{k,t}\!\vbf_i|^{2}\nn\\
& \qquad\qquad\qquad\quad  + \delta^{i}_{k,t} + r_{k}\epsilon^2_{k,t}P + \sigma_{k}^2 \leq 0, \,\, k\in\Kc,i\in\Sc_t, \nn\\
& \quad\,\,\, \|\wbf_t\|^2\leq P.  \nn
\end{align}
By iteratively updating $\vbf$ with the solution $\wbf_t$ to $\Pc^{\text{sca}}_{2,t}(\vbf)$
until convergence, we find a solution to $\Pc^{3\text{fea}}_{2,t}$.

Since problem $\Pc^{\text{sca}}_{2,t}(\vbf)$ is convex,
it can be solved by any existing numerical convex solver.
However, as we need to solve this problem over
all SCA iterations,
it is important to limit the complexity of its solution.
Below, we propose a low-complexity solution based on the
 alternating direction method of multipliers (ADMM) framework \cite{Boyd&etal:book2011}.

The choices of variable substitution can have a substantial impact on the complexity of ADMM. For SegAB, we define the following auxiliary variables:
\begin{align}
& x^{j}_{k,t} \triangleq  \hhbf^{\textsf{H}}_{k,t}\wbf_{j,t}, \quad k\in\Kc,j\in\Sc_t,  \label{eq_x_admm_variable}\\
& \ybf_{i,t} \triangleq  \wbf_{i,t}, \quad i\in\Sc_t.   \label{eq_y_admm_variable}
\end{align}
Also, define
$\xbf_{t} \triangleq [x^{1}_{1,t}, \ldots, x^{S_t}_{K,t}]^{\textsf{T}}\in\mathbb{C}^{S_tK\times 1}$ and
$\ybf_t\triangleq [\ybf_{1,t}, \ldots, \ybf_{S_t,t}]^{\textsf{H}} \in \mathbb{C}^{S_tN\times 1}$.
Problem $\Pc^{\text{sca}}_{2,t}(\vbf)$ can then be transformed to
\begin{align}
& \Pc^{\text{admm}}_{2,t}(\vbf): \, \min_{\xbf_t,\ybf_t,\deltabf_{t},\wbf_t} \, -{\bf{1}}^{\textsf{T}}\deltabf_{t}  \nn\\
& \text{s.t.} \ \ x^{j}_{k,t} = \hhbf^{\textsf{H}}_{k,t}\wbf_{j,t}, \,\, k\in\Kc,j\in\Sc_t, \label{constra_x}\\
& \ybf_{i,t} =  \wbf_{i,t}, \,\, i\in\Sc_t, \label{constra_y}\\
& r_{k}\sum_{j\neq i}|x^{j}_{k,t}|^2 - 2\mu^{i}_{k,t}\Re{\{\vbf^{\textsf{H}}_{i}\hhbf_{k,t}x^{i}_{k,t}\}} + \mu^{i}_{k,t}|\hhbf^{\textsf{H}}_{k,t}\vbf_i|^{2}   \nn\\
& \qquad\qquad + \delta^{i}_{k,t} + r_{k}\epsilon^2_{k,t}P + \sigma_{k}^2  \leq 0, \,\, k\in\Kc,i\in\Sc_t, \label{constra_ADMM_set_f}\\
& \|\ybf_t\|^2 \leq P. \label{constra_ADMM_set_y}
\end{align}

\begin{algorithm}[t]
\caption{Robust Downlink Beamforming Algorithm to
Solve Per-Round Problem $\Pc_{2,t}$}
\label{alg:SegAB}
\begin{algorithmic}[1]
\STATE \textbf{Initialization:} Set arbitrary initial $\mubf_{t}$.
\REPEAT
\STATE \textbf{Initialization:} Set feasible initial $\vbf^{(0)}$. Set $l=0$.
\REPEAT
\STATE \textbf{Initialization:} Set $\wbf^{(0)}_t = \vbf^{(l)}$, $\lambdabf^{(0)}_t = {\bf{0}}$, $\mbf^{(0)}_t = {\bf{0}}$, $n = 0$.
\REPEAT
\STATE Compute $(\xbf^{(n+1)}_t\!,\ybf^{(n+1)}_t\!,\deltabf^{(n+1)}_t)$ by \eqref{x_y_z_update_ADMM}.
\STATE Compute $\wbf^{(n+1)}_{t}$, $\lambdabf^{(n+1)}_t$,
$\mbf^{(n+1)}_t$ by \eqref{delta_w_update_ADMM}--\eqref{m_update_ADMM}.
\STATE Set $n \leftarrow n+1$.
\UNTIL{convergence}
\STATE Set $\vbf^{(l+1)} = \wbf^{(n)}_t$. Set $l\leftarrow l+1$.
\UNTIL{convergence}
\STATE Update $\mubf_{t}$ via \eqref{Update_tau_AO}.
\UNTIL{convergence}
\end{algorithmic}
\end{algorithm}

\begin{figure*}[t]
\centering
\includegraphics[width=0.66\columnwidth]{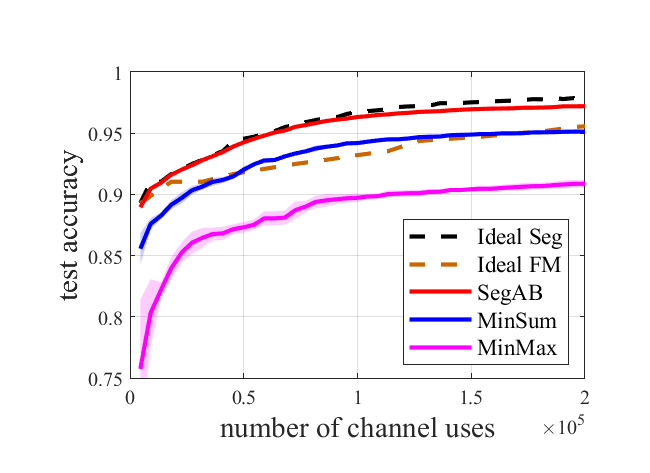}
\includegraphics[width=0.66\columnwidth]{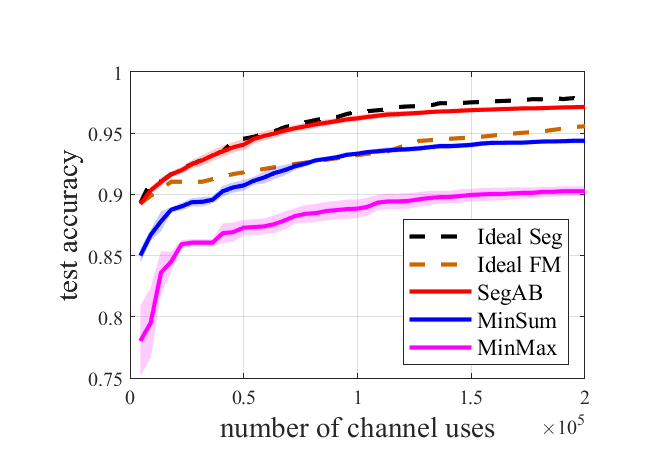}
\includegraphics[width=0.66\columnwidth]{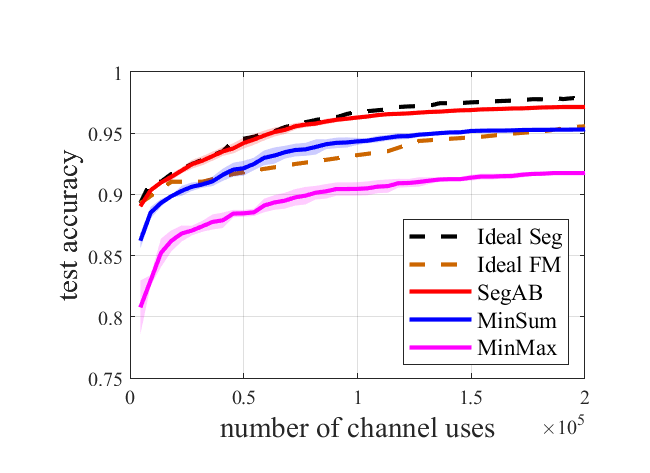}
\vspace*{-1.5em}\caption{Test accuracy vs. number of channel uses for training Model A using MNIST
for $S_t=3,K=5$.
Left: ($(N,\gamma) = (16,0.01)$).
Middle: ($(N,\gamma) = (16,0.1)$).
Right: ($(N,\gamma) = (32,0.1)$).}
\label{Fig2:Acc_MNIST_G3_K5} \vspace*{-1.4em}
\end{figure*}

Let $\Fc_t$ and $\Yc_t$ be the feasible sets for constraints \eqref{constra_ADMM_set_f} and \eqref{constra_ADMM_set_y}, respectively.
Let $I_{\mathcal{F}_t}(\xbf_t,\deltabf_t)$ and $I_{\mathcal{Y}_t}(\ybf_t)$ be the corresponding
indicator functions of these sets.
Let $\lambdabf_t$ and $\mbf_t$ be the dual variables, which are defined similarly to $\xbf_t$ and $\ybf_t$.
We obtain an augmented Lagrangian of $\Pc^{\text{admm}}_{2,t}(\vbf)$,
given by
\begin{align}
& \Lc_{\rho}(\xbf_t,\ybf_t,\deltabf_t,\wbf_t,\lambdabf_t,\mbf_t) =
I_{\mathcal{F}_t}(\xbf_t,\deltabf_t) + I_{\mathcal{Y}_t}(\ybf_t) - {\bf{1}}^{\textsf{T}}\deltabf_{t} \nn\\
&+\frac{\rho}{2}\!\Bigg(\!\sum_{j=1}^{S_t}\!\sum_{k=1}^{K}
\!|x^{j}_{k,t}\!\! -\! \hhbf^{\textsf{H}}_{k,t}\!\wbf_{j,t}
\!\! +\! \lambda^{j}_{k,t}|^{2}\!\! +\! \|\ybf_{t}
\! -\! \wbf_{t}\! +\! \mbf_{t}\|^2\!\!\Bigg)
\label{ADMM_Lagrangian}
\end{align}
where $\rho$ $>$ $0$ is penalty parameter.
Then following the standard ADMM procedure, we decompose $\Lc_{\rho}(\xbf_t,\ybf_t,\deltabf_t,\wbf_t,\lambdabf_t,\mbf_t)$
into two subproblems for $(\xbf_t,\ybf_t,\deltabf_t)$ and $\wbf_t$ separately,
and update them alternatively:
\begin{align}
&\!\! \big(\xbf^{(n+1)}_t\!\!\!,\ybf^{(n+1)}_t\!\!\!,
\deltabf^{(n+1)}_t\!\big)\!\! =\!
\mathop{\arg\!\min}\limits_{\xbf_t,\ybf_t,\deltabf_t}
\!\Lc_{\rho}\big(\!\xbf_t,\!\ybf_t,\!\deltabf_t,
\!\wbf^{(n)}_t\!\!\!,\lambdabf^{(n)}_t\!\!\!, \mbf^{(n)}_t\!\big), \label{x_y_z_update_ADMM}\\
&\!\! \wbf^{(n+1)}_{t}\!\! =\!
\mathop{\arg\!\min}\limits_{\wbf_t}
\!\Lc_{\rho}\big(\!\xbf^{(n+1)}_t\!\!\!,\ybf^{(n+1)}_t\!\!\!,
\deltabf^{(n+1)}_t\!\!\!,\wbf_t,\!\lambdabf^{(n)}_t\!\!\!,
\mbf^{(n)}_t\!\big), \label{delta_w_update_ADMM}\\
&\!\! \lambda_{k,t}^{j(n+1)}\!\! =\! \lambda_{k,t}^{j(n)}
\!\!+\!\big(x_{k,t}^{j(n+1)}\!\!
-\!\hhbf^{\textsf{H}}_{k,t}\!\wbf^{(n+1)}_{j,t}\big),
\,\, k\in\Kc,j\in\Sc_t, \label{lambda_update_ADMM}\\
&\!\! \mbf^{(n+1)}_t\!\! =\! \mbf^{(n)}_t
\!\!+\!\big(\ybf^{(n+1)}_t\!\!
-\!\wbf^{(n+1)}_t\big) \label{m_update_ADMM}
\end{align}
where $n$ is the iteration index.

Note that optimal solutions of the two optimization problems
\eqref{x_y_z_update_ADMM} and \eqref{delta_w_update_ADMM} can be
obtained in a distributed fashion with low-complexity closed-form updates at each ADMM iteration.
We provide the details of the optimal solutions of
\eqref{x_y_z_update_ADMM} and \eqref{delta_w_update_ADMM} in Appendix~\ref{appB}.
Since problem $\Pc^{\text{sca}}_{2,t}(\vbf)$ is convex,
the above ADMM procedure is guaranteed to converge to its optimal solution.

\subsubsection{Algorithm Summary and Properties}
We summarize in Algorithm~\ref{alg:SegAB}
the proposed robust downlink beamforming
algorithm for solving $\Pc_{2,t}$ at each
communication round $t$.

\textbf{Convergence:}
The proposed beamforming algorithm uses ADMM to solve
$\Pc^{\text{sca}}_{2,t}(\vbf)$ optimally
at each SCA iteration, which leads to a solution to problem $\Pc^{3\text{fea}}_{2,t}$.
Following this, by updating $\mubf_t$ via \eqref{Update_tau_AO} based on $\wbf_t$,
our algorithm ensures a monotone decreasing sequence of objective values ${\bf{1}}^{\textsf{T}}\mubf_{t}$,
which are bounded below by zero.
Thus, the algorithm is guaranteed to converge
by the monotone convergence theorem.

\textbf{Computational Complexity:}
The main computational complexity of the proposed
beamforming algorithm is due to
the computation of $\xbf^{(n+1)}_t$ and $\deltabf^{(n+1)}_t$
at each ADMM iteration.
Finding $\xbf^{(n+1)}_t$ and $\deltabf^{(n+1)}_t$
relies on solving $K$ distributed subproblems
$\Pc^{'}_{\text{xdsub}}$ given in Appendix~\ref{appB}
using the interior-point method (IPM)
\cite{Boyd&Vandenberghe:book2004},
each requiring $O(S_t^{3.5}K)$
flops at each ADMM iteration.
The convergence rate of this algorithm
depends on the system parameters $S_t$, $N$, and $K$.
Our experiments show that,
for $S_t=1\sim 20,\forall t$ segments, $N=8\sim 32$ antennas,
and $K=5\sim 20$ devices,
it typical takes $20\sim 300$ ADMM iterations
to converge.

\begin{figure*}[t]
\centering
\includegraphics[width=0.66\columnwidth]{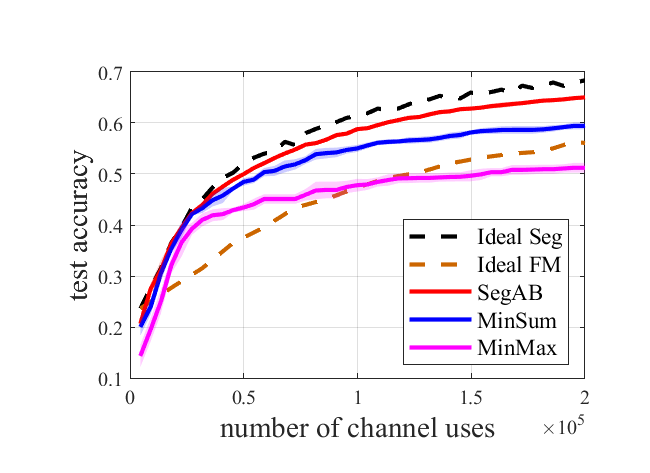}
\includegraphics[width=0.66\columnwidth]{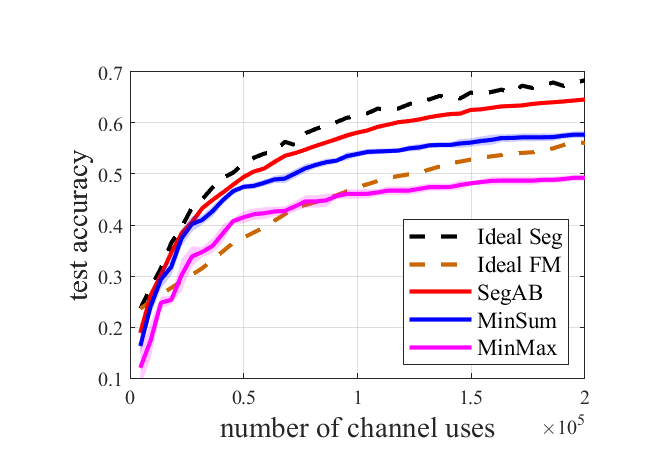}
\includegraphics[width=0.66\columnwidth]{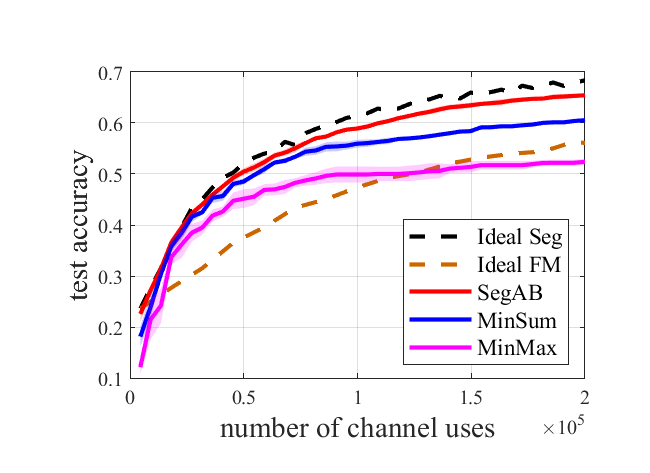}
\vspace*{-1.5em}\caption{Test accuracy vs. number of channel uses for training Model C using CIFAR-10
for $S_t=3,K=5$.
Left: ($(N,\gamma) = (16,0.01)$).
Middle: ($(N,\gamma) = (16,0.1)$).
Right: ($(N,\gamma) = (32,0.1)$).}
\label{Fig3:Acc_Cifar10_G3_K5} \vspace*{-2.2em}
\end{figure*}

\begin{figure}[t]
\centering
\includegraphics[width=0.65\columnwidth]{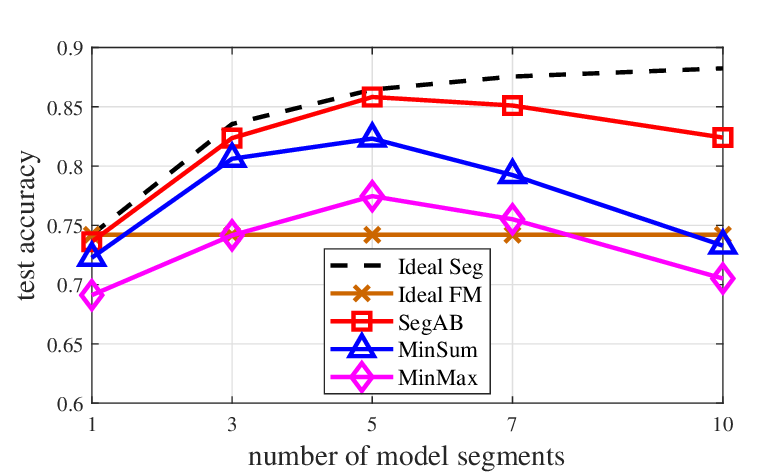}
\vspace*{-.8em}\caption{Test accuracy vs. number of model segments $S_t$ for training Model B using Fashion MNIST
for $(N,\gamma) = (16,0.1)$ and $K=5$.}
\label{Fig4:Acc_FMNIST_N16_gamma01} \vspace*{-1.4em}
\end{figure}

\section{Simulation Results}
\label{sec:simulations}

\subsection{Simulation Setup}
We consider FL training for image classification tasks  under a
cellular  system setting.
Following the typical 5G specifications,
we set system  bandwidth  $10$~MHz and carrier frequency $2$~GHz.
The maximum BS transmit power is $47~\text{dBm}$. Channel vectors are generated i.i.d.\ as $\hbf_{k,t} =
\sqrt{G_{k}}\bar{\hbf}_{k,t}$ with
$\bar{\hbf}_{k,t}\sim\mathcal{CN}({\bf{0}},{\bf{I}})$,
and  $G_k$ being the path gain from the BS to device $k$,\ modeled as
$G_{k} [\text{dB}] = -136.3-35\log_{10}d_k - \psi_k$,
where the BS-device distance  $d_k\in(0.02, 0.5)$  km,
and $\psi_k$ represents shadowing with standard deviation
$8~\text{dB}$.
We set $\epsilon_{k,t} = \gamma\|\hbf_{k,t}\|$
where $\gamma$ is the normalized error bound used for the robust performance study.
The noise power at the device receiver is set to
$\sigma^2 = -96~\text{dBm}$, which accounts for $8~\text{dB}$ noise figure.

We use three different datasets,
MNIST, Fashion MNIST, and CIFAR-10,
for model training and testing.
Each of the MNIST and Fashion MNIST datasets consists of $6\times 10^4$ training samples and $10^4$ test samples from $10$ different classes.
The CIFAR-10 dataset consists of $5\times 10^4$ training samples and
$10^4$ test samples from $10$ different classes.
These datasets are respectively matched with the following
three types of convolutional neural networks:
i) \textbf{Model A}:  an $8\times3\times3$ ReLU convolutional layer,
a $2\times2$ max pooling layer, and a softmax output layer
with $1.361\times 10^4$ parameters.
ii) \textbf{Model B}:  an $8\times3\times3$ ReLU convolutional layer,
a $2\times2$ max pooling layer, a ReLU fully-connected layer with $20$ units, and a softmax output layer
with $2.735\times 10^4$ parameters.
iii) \textbf{Model C}:  a $32\times3\times3$ and two $64\times3\times3$ ReLU convolutional layers (each followed by a $2\times2$ max pooling layer), a ReLU fully-connected layer with $64$ units, and a softmax output layer with $7.3418\times 10^4$ parameters.

The training samples are randomly and evenly distributed  across $K$ devices. For local training using SGD, we set $J=100$, and the mini-batch size $|\Bc^{\tau}_{k,t}|$
is set to $\frac{600}{K}$ for MNIST and Fashion MNIST and to $\frac{500}{K}$
for CIFAR-10, $\forall \tau,k,t$.
We set the learning rate $\eta_t=0.1$, $\forall t$.
For ADMM, we set the penalty parameters $\rho=\tilde{\rho}=0.2$.
All results are obtained by averaging over $5$ drops of device locations, each
with $20$ channel realizations.
We provide $90\%$ confidence interval for each curve, shown as the shadowed area over the curve.

\vspace{-.3em}
\subsection{Performance Comparison}
Besides the proposed SegAB, we consider an optimal benchmark and  three other schemes for comparison:
\begin{mylist}

\item \textbf{Ideal Seg}\cite{Mcmahan&etal:2017}:
Perform segmented downlink broadcast for FL\ via the global model update in \eqref{eq_ul_update_OTA_final},
assuming error-free downlink and
perfect recovery of model parameters at the devices,
\ie receiver noise $\ebf_{k,t} = \bf{0}$.
This   benchmark provides a performance upper bound for all schemes.

\item  \textbf{Ideal FM}: Traditional full-model per-parameter downlink broadcast approach,
assuming error-free downlink and perfect recovery of model parameters at the devices.
It is equivalent to Ideal Seg with $S_t = 1, \forall t$.

\item \textbf{MinSum}: Similar to the proposed SegAB,
except that we obtain the BS transmit beamforming vectors
by replacing $H_t(\wbf_t,\{\epsilon_{k,t}\phibf_{k,t}(\wbf_{t})\})$ in problem $\Pc^{1\text{min}}_{2,t}$
with upper bound $\sum^{S_t}_{i=1}\sum^{K}_{k=1} \frac{ r_k(\sum_{j\neq i}|\hhbf^\textsf{H}_{k,t}\wbf_{j,t}|^2 + \epsilon^2_{k,t}P + \sigma^2) }{ |\hhbf^\textsf{H}_{k,t}\wbf_{i,t}|^2 }$.
We solve the new problem by
projected gradient descent (PGD)
\cite{Mokhtari&etal:NIPS2018},
with a step size $0.01$.

\item \textbf{MinMax}: Similar to the proposed SegAB,
except that we obtain the BS transmit beamforming vectors
by replacing $H_t(\wbf_t,\{\epsilon_{k,t}\phibf_{k,t}(\wbf_{t})\})$ in problem $\Pc^{1\text{min}}_{2,t}$
with upper bound $\max_{ i\in\Sc_t,k\in\Kc } \,
\frac{ r_k(\sum_{j\neq i}|\hhbf^\textsf{H}_{k,t}\wbf_{j,t}|^2 + \epsilon^2_{k,t}P + \sigma^2) }{ |\hhbf^\textsf{H}_{k,t}\wbf_{i,t}|^2  }$.
We solve the new problem by
the projected subgradient algorithm (PSA)
\cite{Zhang&etal:WCL2022},
with a step size $0.01$.

\end{mylist}

We first consider training Model A based on the MNIST dataset.
Fig.~\ref{Fig2:Acc_MNIST_G3_K5} shows the test accuracy performance
vs. the number of channel uses
for different numbers of antennas $N$
and different normalized CSI estimation error
bound values $\gamma$.
We set $S_t=3$ model segments and $K=5$ devices.
We see that SegAB substantially outperforms the other schemes
and nearly obtains the performance of the upper bound Ideal~Seg.
It achieves approximately $97\%$ test accuracy
after $2\times 10^5$ channel uses
for all system configurations $(N,\gamma)$.
Moreover, we see that both MinSum and MinMax converge more slowly than SegAB and achieve lower accuracy.
This demonstrates the advantages of the proposed robust downlink beamforming design.
MinMax has the worst performance among the comparison schemes.
This is because its beamforming design is based on a loose upper bound objective,
which leads to highly suboptimal communication performance
and in turn degrades the learning performance.

\begin{figure}[t]
\centering
\includegraphics[width=0.65\columnwidth]{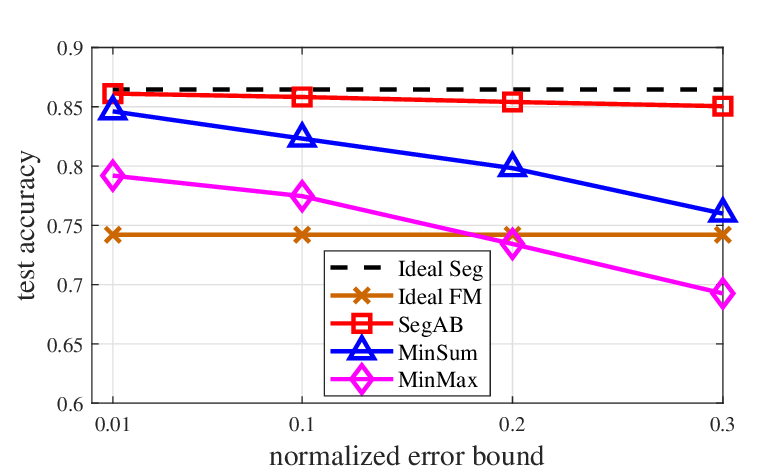}
\vspace*{-.8em}\caption{Test accuracy vs. normalized error bound $\gamma$ for training Model B using Fashion MNIST
for $(N,S_t) = (16,5)$ and $K=5$.}
\label{Fig5:Acc_FMNIST_N16_St5} \vspace*{-1.4em}
\end{figure}

For the more complex dataset CIFAR-10, Fig.~\ref{Fig3:Acc_Cifar10_G3_K5}
shows the test accuracy performance of different methods for training Model C.
We again observe that SegAB
obtains performance close to that of Idal~Seg,
and substantially outperforms the other alternatives
of beamforming for segment broadcast.
Overall, the performance benefit of SegAB is more pronounced with lower channel estimation error. This is expected as they are the conditions for higher beamforming gain.

\subsection{Impact of Other System Parameters}
We now examine the impact of the number of model segments $S_t$
and normalized error bound $\gamma$ on test accuracy.
Figs.~\ref{Fig4:Acc_FMNIST_N16_gamma01} and \ref{Fig5:Acc_FMNIST_N16_St5} compare the test accuracy for training Model B on Fashion MNIST,
under a total of $1\times 10^5$ channel uses at the BS for downlink broadcast.
Fig.~\ref{Fig4:Acc_FMNIST_N16_gamma01} shows the test accuracy
vs.\ $S_t$ for $(N,\gamma) = (16,0.1)$ and $K=5$.
We see that SegAB nearly obtains the performance
of Ideal~Seg for $S_t\leq 5$
and maintains the best performance
among other schemes for all values of $S_t$.
It achieves the best accuracy, approximately $86\%$,
at $S_t=5$ segments.
In contrast, the performance of MinSum and MinMax degrades significantly as $S_t$ increases.

Fig.~\ref{Fig5:Acc_FMNIST_N16_St5} shows the test accuracy vs.\ the normalized error bound $\gamma$
for $(N,S_t) = (16,5)$ and $K=5$.
Again, we see that SegAB nearly attains the optimal performance
under Ideal Seg and outperforms the other alternatives.
The performance gap between SegAB and MinSum or MinMax becomes larger as
$\gamma$ increases. This demonstrates the superior robustness of
the SegAB design.

\section{Conclusion}\label{sec:conclusion}
This paper considers downlink analog broadcast design for
wireless FL under imperfect CSI.
We propose a robust SegAB framework
for sending multiple model segments simultaneously
to reduce communication latency.
Aiming to maximize the FL training performance,
we derive an upper bound on the expected model optimality gap and show that we can minimize it per round via online optimization, independent of the channel states at future rounds.
The per-round downlink beamforming problem is solved via an epigraph-based worst-case optimization, leading to a low-complexity robust method with monotone convergence.
Simulation demonstrates that SegAB
achieves favorable performance against
channel estimation error
and inter-segment interference
for a wide range of system configurations.
It substantially outperforms the conventional full-model
broadcast approach, and other downlink beamforming
alternatives, especially in scenarios with higher number of model segments or higher channel estimation error.

\appendices
\section{Proof of Proposition~\ref{thm:convergence}}\label{appA}
\IEEEproof
We first consider an ideal centralized model training procedure using the full
gradient descent training algorithm and an aggregation of the datasets from all devices.
Let $\vbf^{\tau}_{t}$ be the model update at iteration $\tau$,
with initial point $\vbf^{0}_{t}=\thetabf_{t}$.
The ideal centralized model update is given by
$\vbf^{\tau+1}_{t} = \vbf^{\tau}_{t} - \eta_{t}\nabla F(\vbf^{\tau}_{t})$.
Then, for SegAB, based on \eqref{eq_global_model_update_final},
we have
\begin{align}
& \thetabf_{t+1} - \thetabf^\star = \thetabf_{t} - \eta_{t}\sum^{J-1}_{\tau=0}\nabla F(\vbf^{\tau}_{t}) - \thetabf^\star + \eta_{t}\sum^{J-1}_{\tau=0}\nabla F(\vbf^{\tau}_{t})\nn\\
& + \sum^{K}_{k=1}\frac{A_k}{A}\Delta\thetabf_{k,t} + \sum^{K}_{k=1}\frac{A_k}{A}\ebf_{k,t} = \vbf^{J}_{t}- \thetabf^\star + \alphabf_{t} + \betabf_{t} + \ebf_{t}  \nn
\end{align}
where $\alphabf_{t} \triangleq\eta_{t}\sum^{J-1}_{\tau=0}\big(\nabla F(\vbf^{\tau}_{t})
-\sum^{K}_{k=1}\frac{A_k}{A}\nabla F_{k}(\thetabf^{\tau}_{k,t})\big)$,
$\betabf_{t} \triangleq\eta_t\sum^{K}_{k=1}\frac{A_k}{A}
\sum^{J-1}_{\tau=0}(\nabla F_{k}(\thetabf^{\tau}_{k,t}) - \nabla F_{k}(\thetabf^{\tau}_{k,t}; \Bc^{\tau}_{k,t}))$,
and $\ebf_{t} \triangleq \sum^{K}_{k=1}\frac{A_k}{A}\ebf_{k,t}$.
The term $\alphabf_{t}$ is the aggregated gradient gap
between the ideal centralized
training approach and the wireless FL training approach.
The term $\betabf_{t}$ represents the amount of gradient deviation between using the entire dataset and mini-batches for device local training.
Then, we obtain
\begin{align}
& \mae[\|\thetabf_{t+1} - \thetabf^\star\|^2] = \mae[\|\vbf^{J}_{t}- \thetabf^\star + \alphabf_{t} + \betabf_{t} + \ebf_{t}\|^2] \nn\\
& \leq \mae[(\|\vbf^{J}_{t}- \thetabf^\star\| + \|\alphabf_{t}\| + \|\betabf_{t}\| + \|\ebf_{t}\|)^2] \nn\\
& \stackrel{(a)}{\leq} 4(\mae[\|\vbf^{J}_{t}- \thetabf^\star\|^2]
+ \mae[\|\alphabf_{t}\|^2]
+ \mae[\|\betabf_{t}\|^2]
+\mae[\|\ebf_{t}\|^2])
\nn
\end{align}
where $(a)$ is based on the Cauchy-Schwarz inequality.
We now upper bound each term at the right-hand side of this inequality.
Using the same proof techniques in \cite[Lemma 2]{Bhuyan&etal:2023} and \cite[Lemma 2]{Zhang&etal:2025WiOpt},
under Assumptions~\ref{assump_smooth} and \ref{assump_bound_diff},
for $\eta_t<\frac{1}{L}$, $\forall t\in\Tc$,
we upper bound $\mae[\|\vbf^{J}_{t}- \thetabf^\star\|^2]$ and $\mae[\|\alphabf_{t}\|^2]$, $t\in\Tc$,
as
\begin{align}
& \mae[\|\vbf^{J}_{t}- \thetabf^\star\|^2]
 \leq
(1-\eta_t\lambda)^{2J}\mae[\|\thetabf_{t} - \thetabf^\star\|^2], \label{eq_bound1}\\
& \mae[\|\alphabf_{t}\|^2] \leq \eta^2_{t}J^2\sum^{S_t}_{i=1}\phi_{i}. \label{eq_bound2}
\end{align}
For bounding $\mae[\|\betabf_{t}\|^2]$, we have
\begin{align}
&\mae[\|\betabf_{t}\|^2]  \nn\\
&\! \stackrel{(a)}{\leq}\! \eta^2_{t}\!\sum^{S_t}_{i=1}\! J\sum^{J-1}_{\tau=0}
\mae\!\bigg[\bigg\|
\sum^{K}_{k=1}\!\frac{A_k}{A}\!\Big(\!\nabla F^{i}_{k}(\thetabf^{\tau}_{k,t})\! -\! \nabla F^{i}_{k}(\thetabf^{\tau}_{k,t}; \Bc^{\tau}_{k,t}) \Big)\!\bigg\|^2\bigg] \nn\\
&\! \stackrel{(b)}{\leq}\! \eta^2_{t}\!\sum^{S_t}_{i=1}\! J\sum^{J-1}_{\tau=0}\sum^{K}_{k=1}\!\frac{A_k}{A}
\mae\Big[\Big\|
\nabla F^{i}_{k}(\thetabf^{\tau}_{k,t})\! -\! \nabla F^{i}_{k}(\thetabf^{\tau}_{k,t}; \Bc^{\tau}_{k,t}) \Big\|^2\Big] \nn\\
&\! \stackrel{(c)}{\leq}\! \eta^2_{t}J^2\sum^{S_t}_{i=1}\zeta_{i}
\label{eq_bound4}
\end{align}
where
$(a)$ is based on the Cauchy-Schwarz inequality,
$(b)$ is based on the Jensen’s inequality,
and
$(c)$ follows from Assumption~\ref{assump_bound_diff}.
For bounding $\mae[\|\ebf_{t}\|^2]$, we have
\begin{align}
&\mae[\|\ebf_{t}\|^2] = \sum^{S_t}_{i=1}\mae\bigg[\bigg\|\sum^{K}_{k=1}\frac{A_k}{A}\tilde{\ebf}^{i}_{k,t}\bigg\|^2\bigg] \nn\\
&\!\! \stackrel{(a)}{=} \sum^{S_t}_{i=1}\mae\bigg[\bigg\|
\sum^{K}_{k=1}\frac{A_k}{A}\bigg(\frac{\hhbf^\textsf{H}_{k,t}\wbf_{i,t}\wbf^\textsf{H}_{i,t}\Delta\hbf_{k,t}}{|\hhbf^\textsf{H}_{k,t}\wbf_{i,t}|^2}\tilde{\sbf}_{i,t}
\nn\\
&\!\!+  \sum_{j\neq i}\frac{\hhbf^\textsf{H}_{k,t}\wbf_{i,t}\wbf^\textsf{H}_{j,t}(\hhbf_{k,t} + \Delta\hbf_{k,t})}{|\hhbf^\textsf{H}_{k,t}\wbf_{i,t}|^2}\cdot
\frac{\|\tilde{\sbf}_{i,t}\|\tilde{\sbf}_{j,t}}{\|\tilde{\sbf}_{j,t}\|} +  \tilde{\nbf}^{i}_{k,t}\bigg)\bigg\|^2\bigg] \nn\\
&\!\!  \stackrel{(b)}{\leq} \sum^{S_t}_{i=1}\sum^{K}_{k=1}\frac{A_k}{A}\mae\bigg[\bigg(\frac{|\Delta\hbf^\textsf{H}_{k,t}\wbf_{i,t}|}{|\hhbf^\textsf{H}_{k,t}\wbf_{i,t}|}\|\tilde{\sbf}_{i,t}\|
\nn\\
&\!\!+  \sum_{j\neq i}\frac{|\hhbf^\textsf{H}_{k,t}\wbf_{j,t}| + |\Delta\hbf^\textsf{H}_{k,t}\wbf_{j,t}|}{|\hhbf^\textsf{H}_{k,t}\wbf_{i,t}|}\|\tilde{\sbf}_{i,t}\| +  \|\tilde{\nbf}^{i}_{k,t}\|\bigg)^2\bigg] \nn\\
&\!\!  \stackrel{(c)}{\leq} 4S_t\nu\sum^{S_t}_{i=1}\sum^{K}_{k=1}\frac{A_k}{A}\bigg(\!\frac{ \sum^{S_t}_{j=1} |\Delta\hbf^\textsf{H}_{k,t}\wbf_{j,t}|^2\! +\!\sum_{j\neq i}|\hhbf^\textsf{H}_{k,t}\wbf_{j,t}|^2 }{ |\hhbf^\textsf{H}_{k,t}\wbf_{i,t}|^2  }  \nn\\
&\!\! + \frac{ \sigma^2 }{ |\hhbf^\textsf{H}_{k,t}\wbf_{i,t}|^2  }\!\bigg).
\label{eq_bound5}
\end{align}
where
$(a)$ follows from \eqref{eq_error_equa_accu},
$(b)$ is based on the Jensen’s inequality,
and $(c)$ is based on the Cauchy-Schwarz inequality and the definition of $\nu$.

Combining the above, we have
\begin{align}
& \mae[\|\thetabf_{t+1} - \thetabf^\star\|^2]
\leq G_t\mae[\|\thetabf_{t} - \thetabf^\star\|^2] + H_t(\wbf_t,\Delta\hbf_{t}) + C_{t}.  \nn
\end{align}
Summing up both sides of this inequality over $t\in\Tc$ and  rearranging the terms, we have \eqref{eq_thm1}.
\endIEEEproof

\section{Optimal Solutions to Problems
\eqref{x_y_z_update_ADMM} and \eqref{delta_w_update_ADMM}}\label{appB}

The updates of $\{\xbf^{(n+1)}_t\!,\deltabf^{(n+1)}_t\}$ and $\ybf^{(n+1)}_t$
in problem \eqref{x_y_z_update_ADMM} are independent of each other.
For $\{\xbf^{(n+1)}_t\!,\deltabf^{(n+1)}_t\}$, \eqref{x_y_z_update_ADMM} can be further decomposed into $K$ equivalent smaller problems,
each for $k\in\Kc$, given by
\begin{align}
& \Pc^{'}_{\text{xdsub}}: \,\min_{\{x^{j}_{k,t},\delta^{j}_{k,t}\}^{S_t}_{j=1}} \, \sum_{j=1}^{S_t}\big(|x^{j}_{k,t} - e^{j(n)}_{1,kt}|^{2}
- \frac{2}{\rho}\delta^{j}_{k,t} \big)\nn\\
&\qquad\qquad\quad\text{s.t.} \ r_{k}\sum_{j\neq i}|x^{j}_{k,t}|^2 - 2\mu^{i}_{k,t}\Re{\{e^{i}_{2,kt}x^{i}_{k,t}\}} \nn\\
& \qquad\qquad\qquad\qquad\qquad\qquad + \delta^{i}_{k,t}
+ e^{i}_{3,kt} \leq 0,\,\, i\in\Sc_t \nn
\end{align}
where $ e^{j(n)}_{1,kt} \triangleq \hhbf^{\textsf{H}}_{k,t}\wbf^{(n)}_{j,t} - \lambda^{j(n)}_{k,t}$,
$e^{i}_{2,kt} \triangleq \vbf^{\textsf{H}}_{i}\hhbf_{k,t}$,
and $e^{i}_{3,kt} \triangleq \mu^{i}_{k,t}|\hhbf^{\textsf{H}}_{k,t}\vbf_i|^{2} + r_{k}\epsilon^2_{k,t}P + \sigma_{k}^2$.
Problem $\Pc^{'}_{\text{xdsub}}$ is a convex QCQP, which can be solved optimally by standard convex solvers via IPM \cite{Boyd&Vandenberghe:book2004}.
Based on the KKT conditions, the optimal $\ybf^{(n+1)}_t$ in \eqref{x_y_z_update_ADMM} is given by
$\ybf^{(n+1)}_t = \min{ \Big\{  \frac{\sqrt{P}}{\|\wbf^{(n)}_{t} - \mbf^{(n)}_{t}\|}, 1 \Big\} }(\wbf^{(n)}_{t} - \mbf^{(n)}_{t})$.

Problem \eqref{delta_w_update_ADMM} is an unconstrained convex problem.
The optimal $\wbf^{(n+1)}_{t}$ is given by
$\wbf^{(n+1)}_{j,t}\! =\! ( \sum^{K}_{k=1}\hhbf_{k,t}\hhbf^{\textsf{H}}_{k,t} + \Ibf )^{-1} ( \sum^{K}_{k=1}
( x^{j(n+1)}_{k,t}\! + \lambda^{j(n)}_{k,t} )\hhbf_{k,t} + \ybf^{(n+1)}_{j,t} + \mbf^{(n)}_{j,t} )$, $j\in\Sc_t$.

\balance
\bibliographystyle{IEEEtran}
\bibliography{Refs}

\end{document}